\documentclass[sn-mathphys,Numbered]{sn-jnl}

\usepackage{caption}
\usepackage{subcaption}

\usepackage{amsmath}
\usepackage{array}
\usepackage{booktabs}
\usepackage{graphicx}
\usepackage{threeparttable}
\usepackage{wasysym}
\usepackage{tabularx}
\usepackage{booktabs}
\usepackage{multirow}
\usepackage{makecell}

\usepackage{color, colortbl}

\definecolor{Gray}{gray}{0.9}
\definecolor{Gray2}{gray}{0.8}

\usepackage{graphicx}
\usepackage{multirow}
\usepackage{amsmath,amssymb,amsfonts}
\usepackage{amsthm}
\usepackage{mathrsfs}
\usepackage[title]{appendix}
\usepackage{xcolor}
\usepackage{textcomp}
\usepackage{manyfoot}
\usepackage{booktabs}
\usepackage{algorithm}
\usepackage{algorithmicx}
\usepackage{algpseudocode}
\usepackage{listings}

\theoremstyle{thmstyleone}
\newtheorem{theorem}{Theorem}

\theoremstyle{thmstyletwo}

\theoremstyle{thmstylethree}
\newtheorem{definition}{Definition}

\newcommand{\figsize}{0.73}

\begin{document}

\title[Ariadne]{Ariadne: a Privacy-Preserving Communication Protocol}

\author*[1]{\fnm{Antoine} \sur{Fressancourt}}\email{antoine.fressancourt@huawei.com}

\author[1]{\fnm{Luigi} \sur{Iannone}}\email{luigi.iannone@huawei.com}

\author[2]{\fnm{Mael} \sur{Kerichard}}\email{mael@mael.app}

\affil[1]{\orgname{Huawei Research}, \orgaddress{\city{Paris}, \country{France}}}

\affil[2]{\orgname{ESIR}, \orgaddress{\city{Rennes}, \country{France}}}

\abstract{
    In this article, we present Ariadne, a privacy-preserving communication network layer protocol that uses a source routing approach to avoid relying on trusted third parties.
    In Ariadne, a source node willing to send anonymized network traffic to a destination uses a path consisting in nodes with which it has pre-shared symmetric keys. 
    Temporary keys derived from those pre-shared keys to protect communication privacy using onion routing techniques, ensuring \emph{session unlinkability} for packets following the same path.
  
    Ariadne enhances previous approaches to preserve communication privacy by introducing two novelties. 
    First, the source route is encoded in a fixed size, sequentially encrypted vector of routing information elements, in which the elements' positions in the vector are pseudo-randomly permuted. 
    Second, the temporary keys used to process the packets on the path are referenced using mutually known encrypted patterns. 
    This avoids the use of an explicit key reference that could be used to de-anonymize the communications.
}

\keywords{Privacy, Privacy-enhancing technology, Onion routing}

\maketitle

\section{Introduction}
\label{sec:intro}

Security and privacy have both been afterthoughts in the invention and development of the Internet. 
Yet, while security quickly became a major concern for all Internet users, privacy has only interested specific communities (e.g., researchers, activists, journalists) for a while.
In 2013, the revelation by Edward Snowden of several classified documents shed a new light on privacy in the Internet, making people realize the potential reach of electronic surveillance technologies~\cite{RFC7258}. 
To account for those new concerns, governments have passed laws aiming at protecting users' privacy. 
For instance, in 2016, the European Union parliament adopted the General Data Protection Regulation (GDPR)~\cite{GDPR2016}, which protects European citizens by enforcing strict rules with regards to the storage, exchange, and processing of their personal data. 

To fulfill the need for more privacy, some Internet Service Providers (ISP) started to adopt a ``\emph{privacy by design}'' approach to reduce the amount of personal information they ask their users to expose. 
For instance, in 2020, Google announced that it would soon propose alternatives to third party cookies together with privacy sandboxes in its web browser, Chrome~\cite{Google2020}, 
while Apple launched iCloud Private Relay~\cite{Apple2021}, a proxy-based technology to hide a user's identity from web services providers in June 2021. 
Those commercial solutions follow the trails of research projects and more specialized solutions that aimed at enhancing the privacy of Internet users. 

Some of these projects reuse and enhance the concept of \emph{onion routing}, initially presented in~\cite{goldschlag1996hiding}. 
In onion routing, packets are sequentially encrypted several times, in order to prevent an attacker from observing the path taken by packets in the network.
Those onion routing protocols do either use public key cryptography to protect the communication's privacy at the individual message granularity, like Sphinx~\cite{danezis2009sphinx}, or build private sessions secured using symmetric keys, like Tor~\cite{dingledine2004tor}, in order to adapt to low latency needs.
In a recent work, Kuhn \textit{et al.}~\cite{Kuhn2020}, have revealed a flaw in this framework. They present a packet correlation attack leaving the privacy of sources at risk in front of a global attacker able to corrupt some onion routing nodes. 
This attack has led the authors to propose in the same paper an improvement of the Sphinx protocol.
Later on, the protocol has been also enhanced with the possibility to provide an anonymous return path securely~\cite{Kuhn2021}.
Yet, to this date, and at the best of our knowledge, no proposal has been made to adapt low latency-oriented onion routing protocol to tackle the attack presented in~\cite{Kuhn2020}. 

In this article, we propose a privacy-enhancing network layer protocol that we call Ariadne after the famous ancient Greece myth.\footnote{An early version of this work has been presented at the Cyber Security in Networking conference 2023~\cite{fressancourt23}.}
Ariadne is a source routed, low latency onion routing protocol. 
In Ariadne, a source node can anonymously send packets to a destination along a path consisting in nodes with which it has previously exchanged master symmetric keys.
The communication's privacy is ensured by the use of per-packet symmetric keys derived from the master keys to build the onion layers forming Ariadne packets.
Ariadne can be complemented by a setup onion routing protocol derived from Sphinx~\cite{danezis2009sphinx}, in charge of the exchange of the master symmetric keys with path nodes using Non-Interactive Key Exchanges (NIKE)~\cite{freire2013non}.
We design Ariadne to provide \emph{session unlinkability} against a global passive attacker that can observe network traffic on all the network links, and \emph{path-session unlinkability} against a global attacker able to corrupt some network nodes.
We provide a formal definition of those two unlinkability notions that complement the formal onion security assessment framework proposed in~\cite{Kuhn2020}.

Ariadne encodes the source-routed path in a sequentially encrypted fixed length vector containing a set of routing information elements.
Those elements are placed in the vector according to a pseudo-random permutation. 
This vector and the packet's payload are decrypted on the path to the destination at each intermediate node, preventing packet correlation by external observers. 
The symmetric key to use to process the Ariadne packet at each node is referenced by an encrypted pattern, known by both the source and the intermediate nodes, generated by both nodes using a temporary symmetric key derived from the master key. 

The remaining of the paper is organized as follows. We present the background and related work in Section~\ref{sec:relatedwork}. 
Then, we detail the threat model and the design goals we set for the Ariadne protocol in Section~\ref{sec:threatmodel}. 
We present the main building blocks of the protocol in Section~\ref{sec:ariadnebuildingblocks} and give a formal description of Ariadne in Section~\ref{sec:ariadneproto_desc}. 
The assessment of Ariadne's privacy, and a comparison with state of the art privacy-preserving communication protocols is presented in Section~\ref{sec:priv_properties}.
We present an experimental evaluation of Ariadne in Section~\ref{sec:evaluation}, and introduce future research directions in Section~\ref{sec:ResearchAgenda}. 
Section \ref{sec:conclusion} concludes this article.

\section{Background and Related Work}
\label{sec:relatedwork}

Privacy in communication networks has been addressed in several research projects, whose approaches have been extensively studied in~\cite{shirazi2018survey}. Hereafter we focus on routing and source-destination unlinkability, key management, and privacy protection.

\subsection{Source-Destination Unlinkability}
\label{sec:relatedwork_unlinkability}

Mix networks, introduced by Chaum \textit{et al.}~\cite{chaum1981untraceable}, use a set of nodes to relay a message between a source and a destination, while hiding who are the parties involved in the exchange.
In Sphinx~\cite{danezis2009sphinx}, this protection is provided by the possibility for the sender to describe a path to the destination.
To this end, the source gathers the public keys of the intermediate nodes that the packet will traverse.
Then, it computes a set of shared keys, one for each node, to decrypt the routing information addressed to them.
The data structure describing the path to the destination is sequentially encrypted to prevent intermediate nodes and external observers from accessing information about the entire path.
Topology information is hidden by the fact that this data structure is always of the same size, and intermediate nodes can only read part of it.
The packet relaying operations in Sphinx suffer from a large latency overhead because of the use of public key cryptography.

PHI (Path-Hidden lightweight Anonymity protocol at Network layer~\cite{chen2017phi}) uses a two-step method to address the limitations related to the heavy computing load of public key cryptography in the protection of path information.
In a setup phase, the source sends a packet to a trusted third party mentioning the destination.
While this packet traverses the network, intermediate nodes encrypt the routing information they used to forward the packet and place it in a random order in a vector.
After this phase, the source retrieves the encrypted routing information and uses it in future packets during a data transfer phase.
Similar to PHI, works like TARANET~\cite{chen2018taranet}, Dovetail~\cite{sankey2014dovetail}, and LAP~\cite{hsiao2012lap} rely on some form of trusted third party.
We argue that trusted third party introduce a weakness in the ability of the protocol to preserve the communication's privacy, because the third party might be compromised or pressured to leak information about a target.

HORNET~\cite{chen2015hornet} also uses symmetric keys and a two-step approach to protect the path's privacy, but uses source routing rather than third parties. 
During a setup phase, the source node uses each intermediate node's public key to encrypt routing information they will decrypt when they receive the packet.
This information is a directive telling who the next node is.
On reception of a path setup packet, each intermediate node decrypts the routing directive addressed to it, encrypts it using its (salted) symmetric secret key and adds this privately encrypted routing segment to the setup packet.
The source retrieves the privately encrypted routing segments at the end of the process and uses them to build a source route.
This source route consists in a sequence of privately encrypted segments containing information that can be read only by the intermediate node who has the proper private symmetric key to decrypt it.
We argue that this mechanism suffers from a major privacy issue: even if the private key is salted to alter it from one session to another, the privately encrypted routing information is not changed between two successive packets taking the same path. 
This feature can be used as an attack vector to associate packets belonging to the same flow. 
Indeed, in the article presenting HORNET, the authors mention that this piece of information can be used to perform traffic correlation attacks as presented in~\cite{johnson2013correlation}.

Each of the above-mentioned works has shortcomings due to \emph{(i)} heavy use of public key cryptography, \emph{(ii)} use of third party that can be compromised or \emph{(iii)} possibility to associate packets belonging to the same network flow.

\subsection{Privacy-enhanced Key Management}
\label{sec:relatedwork_crypto}

Privacy-preserving protocols have used various methods to manage the keys used to process packets.

In Sphinx~\cite{danezis2009sphinx}, after each intermediate node has retrieved the routing information, it changes the group element into a value that will be used by the next node, and computes a new routing elements stack.
This procedure allows the protocol to ensure the forward secrecy and packet flow indistinguishability, at the expense of a heavy cryptographic load.

Tor~\cite{dingledine2004tor} uses symmetric key cryptography to protect the privacy of communications between source and destination.
A source sends data to a destination via a set of relay nodes in the form of recursively encrypted cells.
The decryption of each cell is done using a symmetric key identified by a Circuit ID in the cell's header.
As this Circuit ID is similar for all the packets taking the same circuit, it can be used to associate packets to a same data flow.
Leveraging on such a weakness it is possible to carry out a number of attacks compromising the communication's privacy~\cite{basyoni2020traffic}.
HORNET~\cite{chen2015hornet} uses as well symmetric keys, hence, it suffers the same privacy issues as Tor, because the privately encrypted routing information is not changed for two successive packets taking the same path.
Thus, such an information can be used to associate packets belonging to the same flow, easing traffic correlation attacks.

In Anonymous AE~\cite{chan2019anonymous}, Chan and Rogaway try to address the issue of key reference privacy in symmetric key crypto-systems.
To send a packet, the sender takes the content, then adds a number of trailing zeros and encrypts it with one key among a shared key set.
On the receiving end, the receiver retrieves the encrypted content and trailing zeros, and decrypts it using keys picked among the shared key set until it gets a resulting clear text with the appropriate number of trailing zeros.
The receiver has to try several keys before getting the correct clear text, which involves performing a non-deterministic number of useless decryption operations.
As such, this method is not suitable to be used in the operations of a privacy-preserving network protocol, where packet forwarding should be executed as quickly as possible.

\subsection{A Re-Assessment of Onion Routing Privacy}
\label{sec:relatedwork_proof}

For several years, the above-mentioned works have been considered secure enough to protect communication privacy, in a public key setting (Sphinx) or in a low latency context (HORNET, TARANET).
The security assessment of those solutions was based on the formalization of onion routing security made by Camenisch and Lysyanskaya~\cite{Camenisch2005}, in which classic security notions are adapted to give a formal definition of privacy protection levels. 

Kuhn \textit{et al.}~\cite{Kuhn2019} revisited the formal description of the properties associated with privacy-enhancing communication protocols in general, and with onion routing in particular, to better assess the level of protection offered by onion routing protocols. 
The same authors published a follow-up work~\cite{Kuhn2020}, in which the \emph{Garbage attack} is presented. 
This attack consists in altering the payload of a packet sent using an onion routing scheme to allow an attacker controlling several nodes in the network and the receiver of altered packets to correlate packets and deanonymize a packet's source. 
In doing so, an attacker controlling a destination node and a set of onion routing nodes can reduce the anonymity set of the source by making the onion routing nodes alter the payload in a deterministic way, and looking at where the decrypted payload is malformed.
To detect those attacks, Kuhn \textit{et al.} show that onion routing protocols need to protect the integrity of the whole packet between each hop, rather than protecting the header's integrity only.
They give a new formal definition of secure onion routing and present an enhancement to Sphinx to make it secure, but abandon extensions such as replies (present in Sphinx) or sessions (present in low latency onion routing protocols). 
An onion routing protocol with replies has later been presented in~\cite{Kuhn2021}, however, no low latency secure onion routing protocol using sessions (beyond the present paper) has been proposed yet.

\section{Threat Model and Design Objectives}
\label{sec:threatmodel}

Our privacy-preserving protocol aims at protecting against the following aspects of an adversary: 

\makeatletter
\renewenvironment{description}
    {\list{}{
        \leftmargin=10pt
        \labelwidth\z@ \itemindent-\leftmargin
        \let\makelabel\descriptionlabel}}
    {\endlist}
\makeatother

\begin{description}
    \item[\textbf{\emph{Reach:}}] We aim at protecting against a global attacker. 
    In this work, we consider two classes of global attacker: 
    \begin{itemize}
        \item Class $\mathcal{A}_1$ in which the attacker has only access to the links between the network nodes (yet it has access at every link); 
        \item Class $\mathcal{A}_2$ in which the attacker has access to every link in the network and to a set of compromised nodes.
    \end{itemize}
    In our protocol, we assume that at least one of the nodes on the path is not colluding with the others to break the communication's anonymity.
    \item[\textbf{\emph{Activity:}}] In Ariadne, we would like to be able to cope with long-standing passive observers. In case of an active attacker, 
    trying to send forged packets on the network, honest nodes should be able to detect this rogue behavior. We do not consider a DDoS attacker at this stage. 
\end{description}

\noindent To protect against both \textbf{Reach} and \textbf{Activity} of an attacker, we design Ariadne with the following objectives:

\begin{description} 
    \item[\textbf{\emph{Anonymity level:}}] Referring to the terminology presented in~\cite{Kuhn2019}, we aim at ensuring \textbf{\textit{both side unlinkability}}. By observing a packet in the network, an attacker should not be able to determine who are the source and the destination.
    \item[\textbf{\emph{Limited trust:}}] In our protocol, we place only limited trust in other network nodes. In case we consider an adversary from the $\mathcal{A}_2$ class, we assume that at least one node on the path chosen is not colluding with the others. We grant limited trust to the intermediate nodes, so the key material used is temporary.
    \item[\textbf{\emph{Latency and complexity:}}] We want Ariadne to avoid using computationally intensive operations. To do so, we avoid using public key cryptography as much as possible. To address this requirement, like HORNET~\cite{chen2015hornet}, we use symmetric key cryptography to protect the communication's privacy.
    \item[\textbf{\emph{Topological anonymity:}}] If an observer knows the network's topology and routing policy, then knowing the length of the path between source and destination can leak information about source and destination's location. To protect against such topological attacks, we hide the length of the path. Furthermore, intermediate nodes on the path should not be able to determine their position on the path. 
    \item[\textbf{\emph{Forward secure protection:}}] We want to ensure forward secrecy for the packets sent using Ariadne. This requires changing encryption keys with each packet. 
    \item[\textbf{\emph{Session unlinkability:}}] To address the challenges raised in~\cite{Kuhn2020}, we want to design a privacy preserving protocol that is secure against the Garbage attacks, also providing session functionality, \textit{i.e.}, packets can be sent from a source to a destination following the same path while packets belonging to the same session cannot be linked together. By preventing such a correlation, we want to protect against traffic correlation attacks (as presented in~\cite{johnson2013correlation}). 
\end{description}

\section{The building blocks used in Ariadne}
\label{sec:ariadnebuildingblocks}

Ariadne is a source routed onion protocol and its novelty, compared with previous privacy-preserving protocols, is based on two building blocks: \emph{(i)} the routing elements vector format that is used to encode the route taken by the packet; and \emph{(ii)} the method used to reference the temporary symmetric keys used to decrypt a packet. 
In the following we first describe how Ariadne encodes the source route in a \emph{routing information vector}, then we show how symmetric keys are derived and referenced.

\subsection{Ariadne's Routing Information Vector}
\label{sec:ariadneproto_pathencoding}

The routing information to be used by each node on the path is sequentially encrypted by the source using temporary keys derived from a set of pre-shared master symmetric keys.
Each path node can retrieve the data structure encoding the path, decrypt it to access the routing information it needs (and only this information), and re-encrypt it to blind this information for the other nodes. 

To carry this information, we use a fixed size vector, in which the location of the routing information elements are permuted randomly, to avoid intermediate nodes to be able to determine their position in the path from their slot in the vector.
In Sphinx~\cite{danezis2009sphinx}, the routing information is stored in a sequentially encrypted Last in, First out stack. 
One of the drawbacks of such a data structure is that nodes processing a Sphinx packet need to completely rewrite large parts of the packet header before relaying it to the next node.
Processing the routing information vector only requires doing exclusive OR operations with pseudo-random byte strings, which is better from a performance perspective.  

\subsection{Ariadne's Symmetric Keys Derivation and Reference}
\label{sec:ariadneproto_keyderivation}

Passing a key reference to decrypt information in a privacy preserving protocol is a major issue as a key identifier can be used to associate packets together and thus harm the privacy of the communication.
While in theory public key cryptography offers tools to negotiate per packet symmetric keys between a source and various intermediate nodes in a non-interactive way, this method is too heavy from a computing perspective to be used in low latency contexts.
The use of pre-computed shared symmetric keys can be beneficial in that regard, but then specific care must be taken to prevent the key identifier from allowing to identify the flow packets belong to.

To avoid such a problem, secure communication protocols, like Ariadne, can benefit from key derivation methods as they can be used to compute numerous keys from a single master key, reducing the communication load due to  key negotiation. For instance, to prevent attacks related to the reuse of symmetric keys, credit card payment systems use the Derive Unique Key Per Transaction (DUKPT) scheme~(\cite{ansiDUKPT}, \cite{brier2010forward}) to compute per transaction symmetric keys from a master key provisioned in each credit card's secure element. 
In other contexts, key derivation from a master key can be done using HMAC-based Key Derivation Functions (HKDF)~\cite{krawczyk2010HKDF}. 
As both DUKPT and HKDF use symmetric key cryptography, they also reduce the computing load related to key derivation operations. 

\begin{figure}[t]
    \centering
    \includegraphics[width=.90\linewidth]{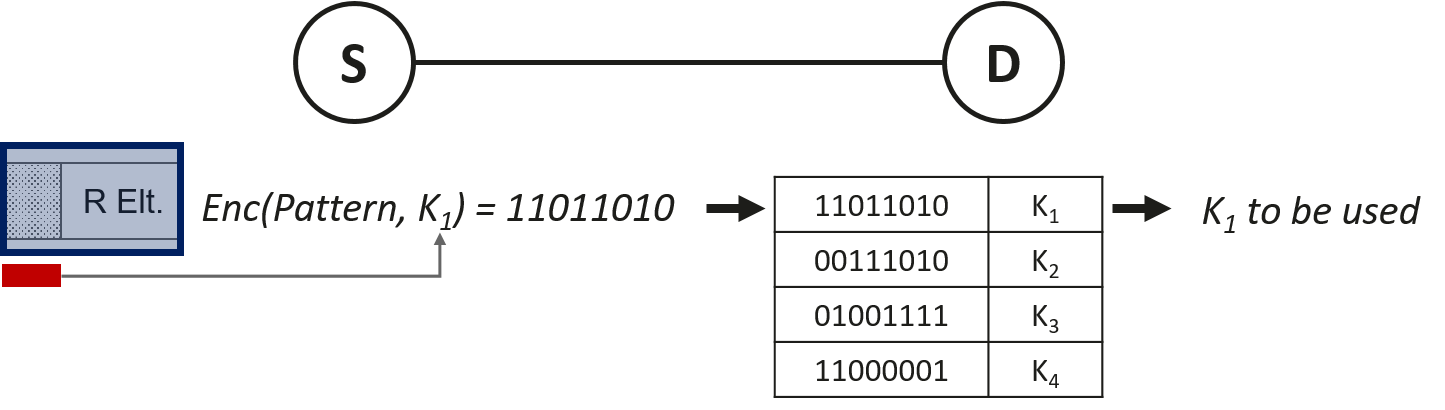}
    \caption{Referencing a shared key using an encrypted pattern}
    \label{fig:sub-KeyReference}
\end{figure}

The intermediate nodes can compute a set of temporary keys using a predefined key derivation mechanism and encrypt a predefined pattern using those keys. 
Then they can place the keys and the encrypted patterns in a matching table. 
When the source wants to send routing information to the intermediate node, it can prepend this information with the predefined pattern and encrypt the resulting information block with a key derived from a shared master key using the same key derivation mechanism as the intermediate nodes. 
When the intermediate node receives the encrypted routing information element, it can identify the key it must use by looking up for the encrypted pattern in the table it previously built (see Figure \ref{fig:sub-KeyReference}).
This method trades off a bit of memory at intermediate nodes for the sake of a quicker packet processing.

\section{Ariadne's Formal Description}
\label{sec:ariadneproto_desc}

In this section, we formally describe the creation and processing of Ariadne packets along a path. We assume that the source node has a shared master symmetric key with each node on the path. We consider a path from a source node \textbf{$N_0$} to a destination node \textbf{$N_{n+1}$}, going through a set of intermediate nodes \textbf{$N_1$}, \textbf{$N_2$}... and \textbf{$N_{n}$}. 

As explained earlier, we assume that $N_0$ has negotiated or retrieved a set of shared master keys \textbf{$k_1$}, \textbf{$k_2$}... \textbf{$k_n$} and \textbf{$k_{n+1}$} with the nodes along the path to the destination \textbf{$N_{n+1}$}. The required key exchanges can be done following a similar method as in TARANET~\cite{chen2018taranet}, where an enhanced version of the Sphinx protocol is used by the source to retrieve the forwarding segments created by the intermediate nodes to encode the routing information they need. An example of such solution can be found in~\ref{sec:ariadneproto_desc_setup}.
Other key exchange and distribution mechanisms could be used depending on the context in which Ariadne operates. For instance, the control plane used by the nodes to determine the paths to reach other nodes could distribute the shared symmetric keys alongside the source route to a destination. To avoid a privacy leakage associated with this distribution, concepts borrowed from Private Information Retrieval~\cite{kushilevitz1997pir} could be used at a benefit.

Further, all the nodes have agreed on a common byte pattern $\pi$ used to reference the temporary key to use to process the Ariadne packet.
Each node on the path encrypts this pattern using temporary keys derived using a deterministic and mutually agreed upon key derivation method, and place the resulting encrypted pattern in a matching table, following the method presented in Section~\ref{sec:ariadneproto_keyderivation}, to determine quickly the temporary symmetric key to use to decrypt the routing information elements.

The packet creation and relaying operations described in this section only use symmetric key cryptographic operations and matching actions against tables.
In the rest of this section, after introducing the notation (Section~\ref{sec:ariadneproto_data_notations}), we will present how Ariadne packet are created by the source \textbf{$N_0$} (Section~\ref{sec:ariadneproto_data_creation}).

\subsection{Notation}
\label{sec:ariadneproto_data_notations}

We denote $\{0 ; 1\}_{a}$ a random bit string of length $a$, in which bits are randomly set to $0$ or $1$. 
We denote $0_a$ a string of $0$ bits of length $a$. 
We denote $x_{[a; b]}$ the substring of $x$ taken from bit $a$ to bit $b$ inclusive. We interchangeably use the notation $|x|$ to express the length in bits of the string $x$ or to convert a variable expressing a length in bytes to a length in bits. 
We denote $x|y$ the concatenation of strings $x$ and $y$.

In Ariadne, a \emph{data routing elements vector} is used to convey the anonymized source-routed path information, denoted as $R$. 
The data routing elements of the vector are denoted by $r_{i,t}$  and used by node $N_i$ to route the packet indexed by variable $t$ (where $t$ is incremented for each packet following the same path).
For $i$ ranging from $1$ to $n$, this routing element is the concatenation of four elements: $\pi$, a byte pattern indexing the key as explained in Section~\ref{sec:ariadneproto_keyderivation}; $Addr_{i+1}$, the address of $N_i$'s successor on the path; $p_{{i+1},t}$, the pointer giving the position of the next routing element in the vector; and $\gamma_{i,t}$, the result of a MAC (Message Authentication Code) computed over the whole packet (including the header).
In $r_{n+1,t}$, as there is no successor node, the address is $N_{n+1}$'s own address and the pointer is pointing to the actual slot in the routing element vector.
This routing element is positioned at byte indexed by $P_{i,t}$, and has a length of $U$ bytes.

Let us denote with $l_{pmax}$ the maximum number of relays that can be accommodated by the routing element vector; then the data routing element vector's length is $L_{dmax} = l_{pmax} \times U$ bytes.
We denote $X$ the packet's content byte stream (header and payload) being created by the source and processed on the path.
The packet including the payload is $L_{full}$ bytes long. 
If necessary, the payload is padded to maintain a fixed packet length.

Finally, let us define the following cryptographic elements of the protocol:

\begin{itemize}
    \item $\kappa$ is the length of the shared secret keys used by the nodes.
    \item $k_i$ denotes the shared master symmetric key negotiated by $N_0$ with $N_i$, while $k^{Enc}_{i, t}$ and $k^{Mac}_{i, t}$ are two temporary keys derived from $k_i$ and indexed by the parameter $t$ that are used respectively to encrypt and to check the integrity of a packet.
    \item $\delta: \{0,1\}^\kappa \times \mathbb{N} \rightarrow \{0,1\}^\kappa \times \{0,1\}^\kappa$ is a key derivation function based on the O-DUKPT method allowing to compute the two temporary keys $k^{Enc}_{i, t}$ and $k^{Mac}_{i, t}$ from a master key $k_i$ and the parameter $t$.
    \item  $\rho: \{0,1\}^\kappa \rightarrow \{0,1\}^{L_{dmax}+U_d}$ is a pseudo-random generator keyed by $k^{Enc}_{i, t}$ to generate a string of length $L_{dmax}+U_d$ bytes.
    \item  $\mu: \{0,1\}^\kappa \times \{0,1\}^{L_{dmax}+U} \rightarrow \{0,1\}^\kappa$ is a Message Authentication Code (MAC) function over the whole packet. It is keyed by $k^{Mac}_{i, t}$ to generate a MAC $\gamma$ of length $\kappa$ bytes.
\end{itemize}

Following the formal definition of onion routing protocols presented in~\cite{Kuhn2020}, we will formally describe the Ariadne protocol as an Onion Routing scheme consisting in three algorithms: a key generation algorithm; an onion creation algorithm used by the source to form the packet to send on the network (see Section~\ref{sec:ariadneproto_data_creation}) and an onion processing algorithm used by nodes $N_1$...$N_{n+1}$ to relay the packet on the path (see Section~\ref{sec:ariadneproto_data_processing}).

\subsection{Ariadne packet creation}
\label{sec:ariadneproto_data_creation}

Ariadne packet creation is done in two steps, namely a key generation step and an onion creation step.

\noindent\textbf{Key generation algorithm:} 
~\\
\textbf{$N_0$} gathers the list of nodes \textbf{$N_1$}, ... \textbf{$N_n$} and \textbf{$N_{n+1}$} on the path to the destination, as well as the symmetric master keys it shares with them $k_0$, $k_1$, ... $k_{n}$ and $k_{n+1}$.
Then, $N_0$ computes the cryptographic material for creating the packet indexed by $t$ as follows:
\begin{equation}
    \begin{aligned}
        &(k^{Enc}_{i, t}; k^{Mac}_{i, t}) = \delta(k_i, t)\\
        &\rho_{i,t,\pi} = \rho(k^{Enc}_{i, t})_{[0; |\pi|]}\\
        &\rho_{i,t,r} = \rho(k^{Enc}_{i, t})_{[0; |U-1|]}\\
        &\rho_{i,t,Header} = \rho(k^{Enc}_{i, t})_{[|U|; |U + L_{dmax}|]}\\
        &\rho_{i,t,Packet} = \rho(k^{Enc}_{i, t})_{[|U|; |L_{full}-1|]}
    \end{aligned}
\end{equation}\\

\begin{figure}[t]
    \centering
    \begin{subfigure}{\figsize\textwidth}
        \includegraphics[width=\linewidth]{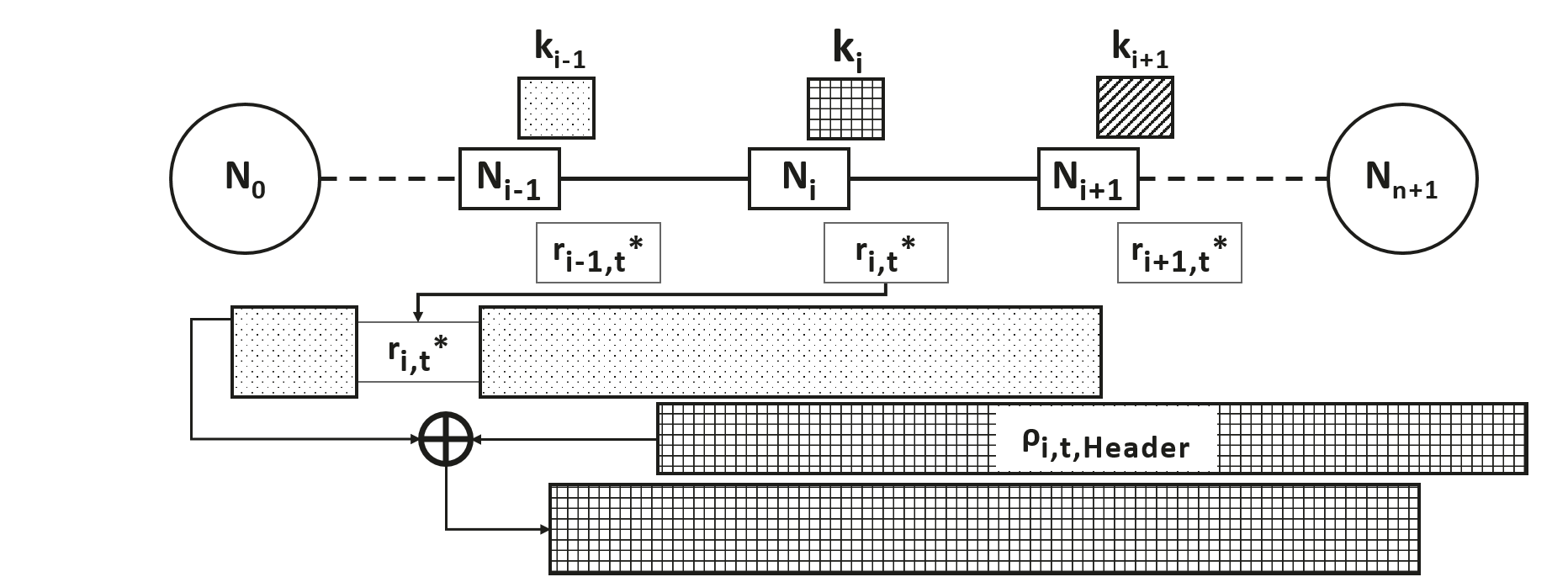}
        \captionsetup{justification=centering}
        \caption{Preparing the routing \\ element filler string}
        \label{fig:sub-PacketCreation01}
    \end{subfigure}
    \medskip
    \begin{subfigure}{\figsize\textwidth}
        \includegraphics[width=\linewidth]{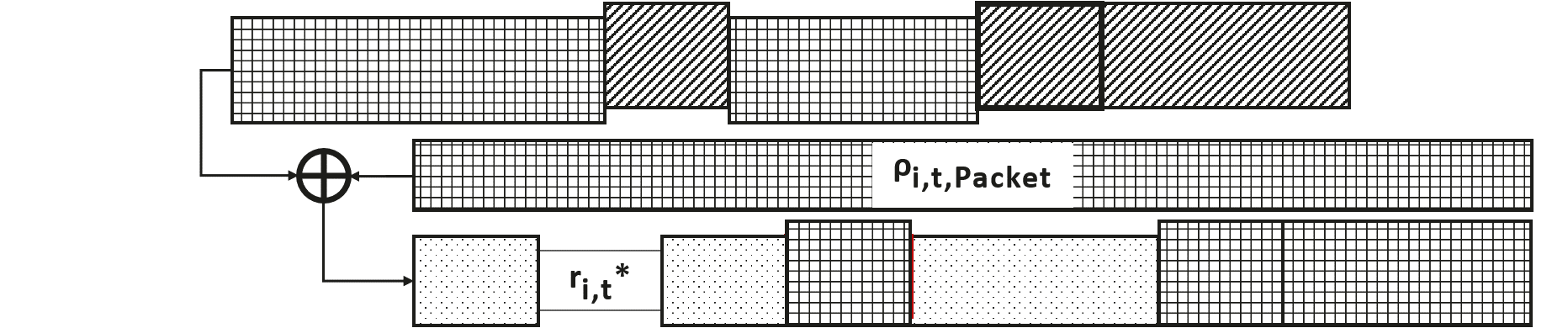}
        \captionsetup{justification=centering}  
        \caption{The packet and routing element are \\ encrypted using $\rho_{i,t,Packet}$}
        \label{fig:sub-PacketCreation02}
    \end{subfigure}
    \medskip
    \begin{subfigure}{\figsize\textwidth}
        \includegraphics[width=\linewidth]{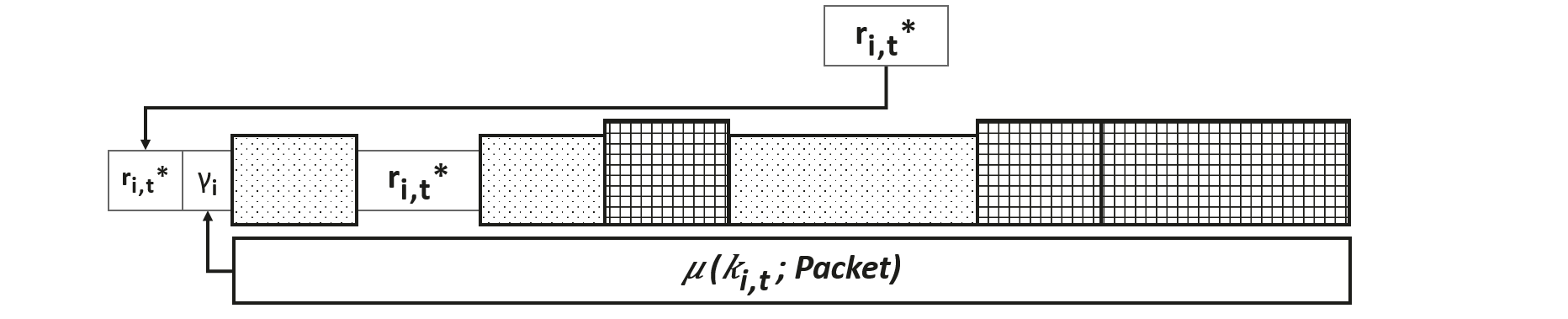} 
        \captionsetup{justification=centering} 
        \caption{A MAC is computed for the \\ routing element vector and payload}
        \label{fig:sub-PacketCreation03}
    \end{subfigure}
    \medskip
    \begin{subfigure}{\figsize\textwidth}
        \includegraphics[width=\linewidth]{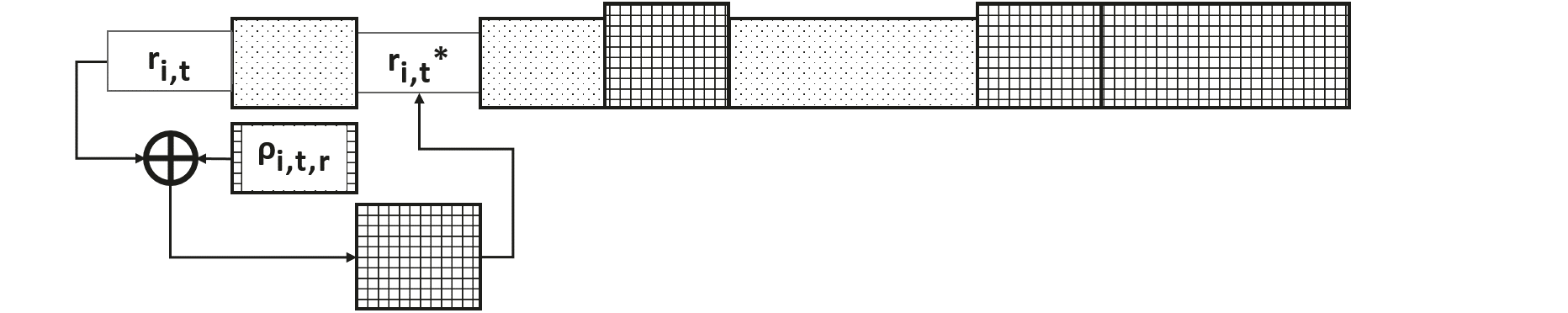} 
        \captionsetup{justification=centering} 
        \caption{$r_{i,t}$ is encrypted using $\rho_{i,t,r}$ \\ and placed at slot $p_{i,t}$}
        \label{fig:sub-PacketCreation04}
    \end{subfigure}
    \caption{Packet creation procedure: steps performed for node $N_i$}
    \label{fig:PacketCreation}
\end{figure}

\noindent \textbf{Onion creation algorithm:} 
~\\
After computing the cryptographic material, the source \textbf{$N_0$} initializes a routing elements vector consisting of $l_{pmax}$ slots of size $U$ bytes. 
\begin{equation}
    \begin{aligned}
        &R = \{0 ; 1\}_{|L_{dmax}|}\\
    \end{aligned}
\end{equation}
Then, $N_0$ computes a pseudo permutation over the $l_{pmax}$ slots in the routing elements vector to determine the slot positions $p_{1,t}$, ... $p_{n,t}$ and $p_{n+1,t}$ at which it will place the routing information elements that will be used by each node. 
During the onion formation, $N_0$ will have to compute a set of MACs over the routing element vector and payload of the packet.
To do so, it builds a filler bit string using the $r^*_{i,t}$ routing element (see Figure~\ref{fig:sub-PacketCreation01}).
Starting from $N_1$'s information until $N_{n+1}$'s, $N_0$ build $R$ from the initial random bit string by doing the following operations (described for node $N_i$):
\begin{equation}
    \begin{aligned}
        &r^*_{i,t} = \pi | Addr_{i+1} | p_{i+1,t} | 0_\kappa\\
        &R = R_{[0; |P_{d,i,t}-1|]} | r^*_{i,t} | R_{[|P_{d,i,t} + U|; |L_{dmax}-1|]}\\
        &R = R \oplus \rho_{i,t,Header}\\
    \end{aligned}
\end{equation}
Once the filler string for $R$ is prepared, $N_0$ can start building the packet, as presented in Figure~\ref{fig:sub-PacketCreation02}. 
This construction is made backwards, starting from the destination $N_{n+1}$.

First, $N_0$ appends the payload to $R$. If the payload is shorter than a fixed length $L_{Payload}$, it is padded to fit this size in order to prevent an attacker from using the packet's length to de-anonymize the communication.
Then, for each node on the path, starting with node $N_{n+1}$ down to node $N_1$, $N_0$ follows the same algorithm (described for node $N_i$).
The source node $N_0$ encrypts the packet using the $\rho_{i,t,Packet}$ string. It computes the MAC $\gamma_{i,t}$ of the resulting packet using $\mu$ keyed by $k^{Mac}_{i, t}$ (Figure~\ref{fig:sub-PacketCreation03}). It uses it together with the pattern $\pi$, the address of $i$'s successor on the path and the position of this successor's routing element in the vector to compute $r_{i,t}$. After being encrypted using $\rho_{i,t,r}$, the routing element is placed in the vector $R$ (Figure~\ref{fig:sub-PacketCreation04}). The formal definition of such operation is:
\begin{equation}
    \begin{aligned}
        &X = X \oplus \rho_{i,t,Packet}\\
        &\gamma_{i,t} = \mu(k^{Mac}_{i, t}; X)\\
        &r_{i,t} = \pi | Addr_{i+1} | p_{i+1,t} | \gamma_{i,t}\\
        &X = X_{[0; |P_{i,t}-1|]} | r_{i,t}  \oplus \rho_{i,t,r} | X_{[|P_{i,t} + U|; |L_{full}|]}\\
    \end{aligned}
\end{equation}
At the end of the routing vector creation procedure, a common header containing $N_1$'s address $Addr_{N_1}$ is prepended in clear and the pointer $p_{1,t}$ to the routing vector element that $N_1$ has to process to find out the next hop. The packet is then ready to be sent to $N_1$.

\subsection{Ariadne packet processing}
\label{sec:ariadneproto_data_processing}

\begin{figure}[t]
    \centering
    \begin{subfigure}{\figsize\textwidth}
        \includegraphics[width=\linewidth]{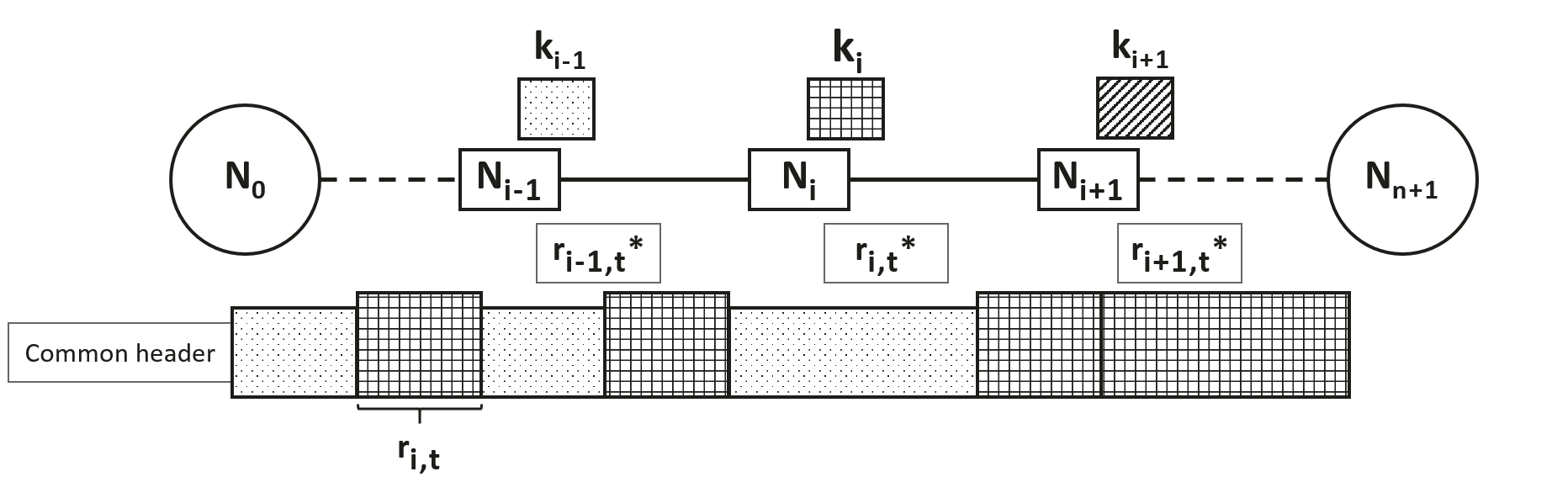}
        \captionsetup{justification=centering}
        \caption{Packet received at node $N_i$. \\ $N_i$ retrieves $k^{Enc}_{i, t}$ from the lookup of $\pi_{Enc}$.}
        \label{fig:sub-PacketRelay01}
    \end{subfigure}
    \medskip
    \begin{subfigure}{\figsize\textwidth}
        \includegraphics[width=\linewidth]{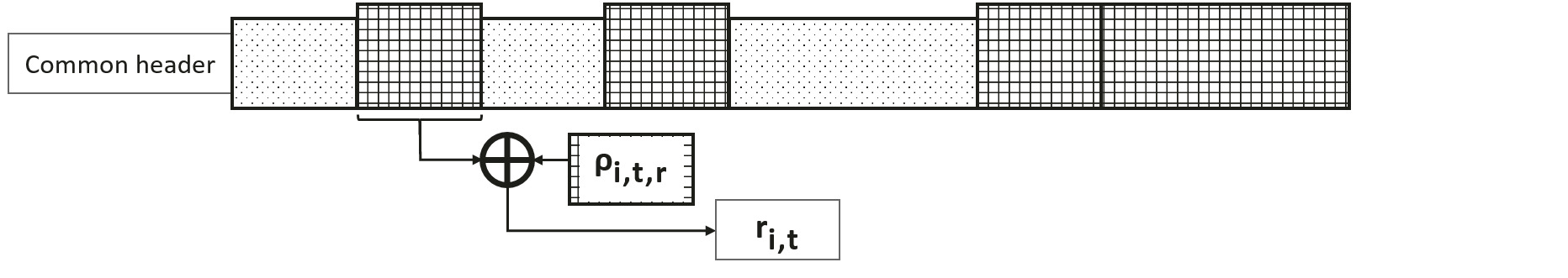}
        \captionsetup{justification=centering}  
        \caption{Retrieval of $r_{i,t}$ from the routing element vector}
        \label{fig:sub-PacketRelay02}
    \end{subfigure}
    \medskip
    \begin{subfigure}{\figsize\textwidth}
        \includegraphics[width=\linewidth]{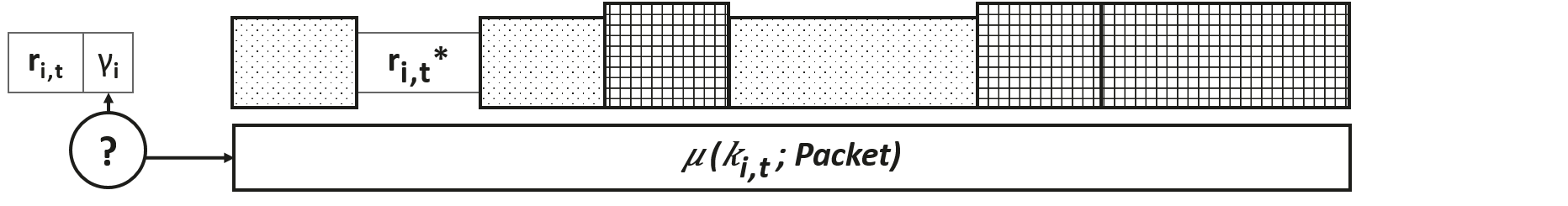} 
        \captionsetup{justification=centering} 
        \caption{Computation of $\gamma^{\prime}_{i,t}$ and comparison to $\gamma_{i,t}$}
        \label{fig:sub-PacketRelay03}
    \end{subfigure}
    \medskip
    \begin{subfigure}{\figsize\textwidth}
        \includegraphics[width=\linewidth]{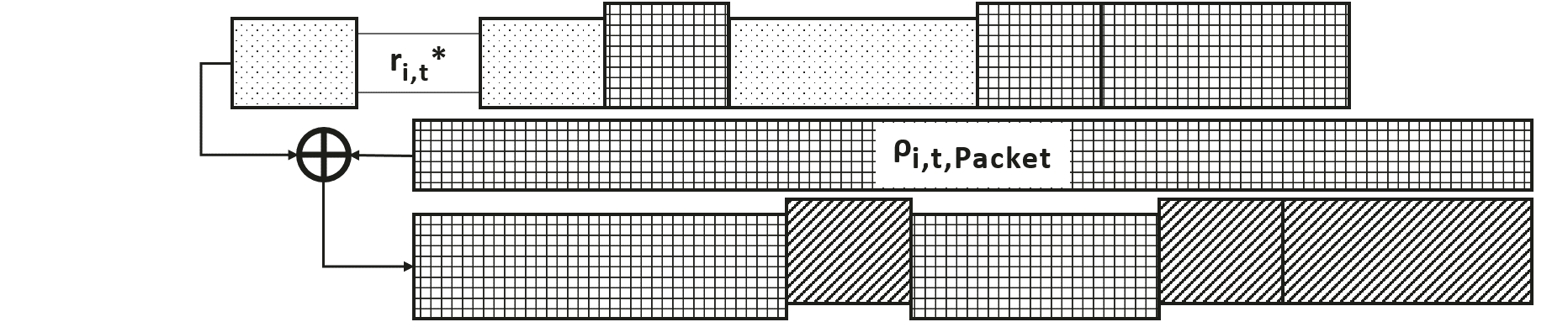} 
        \captionsetup{justification=centering} 
        \caption{Decryption of the packet and blinding of $r_{i,t}$}
        \label{fig:sub-PacketRelay04}
    \end{subfigure}
    \medskip
    \begin{subfigure}{\figsize\textwidth}
        \includegraphics[width=\linewidth]{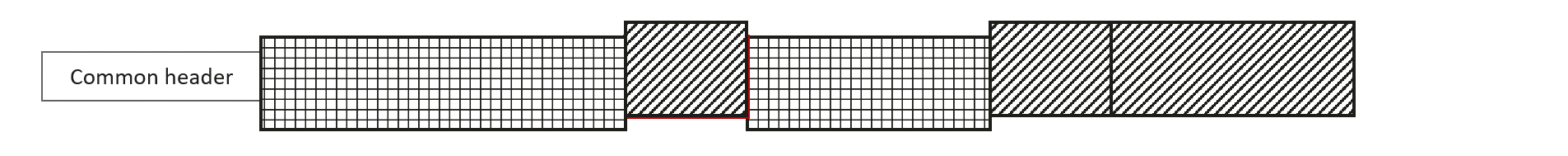} 
        \captionsetup{justification=centering} 
        \caption{$N_i$ updates the common header}
        \label{fig:sub-PacketRelay05}
    \end{subfigure}
        
    \caption{Packet relaying procedure at node $N_i$}
    \label{fig:PacketProcessing}
\end{figure}

We describe the packet processing operations (following the \textbf{Onion processing algorithm}) performed by a node $N_i$ on the path between $N_0$ and $N_{n+1}$ (see Figure~\ref{fig:PacketProcessing}). 

When $N_i$ receives the packet (Figure~\ref{fig:sub-PacketRelay01}), it retrieves the routing element it must process by looking at the pointer $p_{i,t}$ in the common header. Then it takes the beginning of the routing element and looks for a match with the encrypted patterns stored in the pattern matching table it built to retrieve the temporary shared symmetric keys to use to process the packet, ($k^{Enc}_{i, t}$ and $k^{Mac}_{i, t}$).
If a match is found, it derives the cryptographic material to use to process the packet:
\begin{equation}
    \begin{aligned}
        &\pi^{\prime}_{Enc} = X_{[|P_{i,t}|; |P_{i,t}| + |\pi|]}\\
        &\pi^{\prime}_{Enc} \overset{?}{=} \pi \oplus \rho_{i,t,\pi} \rightarrow (k^{Enc}_{i, t}; k^{Mac}_{i, t})\\
        &\rho(k^{Enc}_{i, t}) \rightarrow (\rho_{i,t,r}; \rho_{i,t,Header}; \rho_{i,t,Packet})\\
    \end{aligned}
\end{equation}
$N_i$ then decrypts the routing element with $\rho_{i,t,r}$ to obtain $r_{i,t}$ (Figure~\ref{fig:sub-PacketRelay02}). Then it looks at the decrypted routing element and retrieves $\gamma_{i,t}$, the MAC for the packet.
\begin{equation}
    \begin{aligned}
        &r_{i,t} = X_{[|P_{i,t}|; |P_{i,t} + U - 1|]} \oplus \rho_{i,t,r}\\
        &r_{i,t} \rightarrow (r^*_{i,t}; \gamma_{i,t})\\
    \end{aligned}
\end{equation}
It replaces $r_{i,t}$ with $r^*_{i,t}$ to compute  $\gamma^{\prime}_{i,t}$, the MAC of the packet it received. It is compared to $\gamma_{i,t}$ and if they are the same the packet is further processed (cf. Figure~\ref{fig:sub-PacketRelay03}). Otherwise, the packet is dropped to avoid being used to carry out a Garbage attack. 
\begin{equation}
    \begin{aligned}
        &X = X_{[0; |P_{i,t}-1|]} | r^*_{i,t} | X_{[|P_{i,t} + U|; |L_{full}|]}\\
        &\gamma^{\prime}_{i,t} = \mu(k^{Mac}_{i, t}; X)\\
        &\gamma^{\prime}_{i,t} \overset{?}{=} \gamma_{i,t} \rightarrow X = X \oplus \rho_{i,t,Packet}\\
    \end{aligned}
\end{equation}

The packet is decrypted using $\rho_{i,t,Packet}$ and the address and slot pointers retrieved from $r_{i,t}$ are placed in the common header (Figure~\ref{fig:sub-PacketRelay04}).
\begin{equation}
    \begin{aligned}
        &X = X \oplus \rho_{i,t,Packet}\\
    \end{aligned}
\end{equation}
This decryption of the packet using $\rho_{i,t,Packet}$ blinds at the same time the routing information element $r_{i,t}$. The packet is then sent to the next node, $N_{i+1}$. When $N_{n+1}$ receives the packet, it begins processing it like the previous intermediate nodes. When it finds out the address of the next node in $r_{n+1,t}$ is its own address, then it checks the packet integrity, and if the packet has not been corrupted it proceeds with the decryption of the packet. 

\section{Ariadne's Privacy Properties}
\label{sec:priv_properties}

One of the main objectives of the Ariadne protocol is to add a notion of session to secure onion routing protocols. To formally define this notion, we need to extend the three secure onion routing properties defined in \cite{Kuhn2020} (Onion correctness, Tail-Indistinguishability, and Layer-Unlinkability). Hereafter we provide a formal definition of \emph{session unlinkability}, followed by an assessment regarding how our protocol provides this property in our threat model.    

\subsection{Unlikability Definitions}
\label{sec:unlinkability_def}

In order to give a formal definition of session unlinkability, we define two weaker properties: \emph{path-session unlinkability} and \emph{source-session unlinkability}.

\begin{description}
\item[\textbf{Path-session unlinkability}] means that an adversary cannot determine if two packets originating from the same source and passing through the same honest node $N_j$ belong to the same session, \textit{i.e.}, if they have taken the same path before node $N_j$.

\begin{definition}[\textbf{Path-session unlinkability}]
\label{def:path-session-unlinkability}

Consider an adversary interacting with a challenger running our privacy-preserving protocol. The path session unlinkability game runs as follows:
  
\begin{enumerate}
    \item [1.] The adversary receives as input the name of a honest router $N_j$.
    \item [2.]  The adversary may submit any number of packets to the challenger, which returns the result of the processing of those packets by $N_j$.
    \item [3.] The adversary submits a pair of packet payloads $(p_1; p_2)$ and a path $\mathcal{P}=(N_0, N_1, ..., N_j, ... N_{n+1})$, in which $N_j$ occupies the $j^{th}$ position in the path. If its adversary class allows it, the adversary also provides the shared symmetric master keys $k_i$ shared between the source $N_0$ and all the nodes $N_i$ in the network, except $N_j$. 
    \item [4.] The challenger sets up shared symmetric master keys between $N_0$ and every other node $N_i$ (if they have not been provided by the attacker).
    \item [5.] The challenger sets a bit $b$ at random. It creates a first data packet ($Pkt_1$) carrying $p_1$ following path $\mathcal{P}$. Then it creates two other packets carrying $p_2$. A first packet ($Pkt_2$) is following path $\mathcal{P}$. A second packet ($Pkt_2^{\prime}$) is following path $\mathcal{P}^{\prime}=(N_0, N_1^{\prime}, ..., N_{k-1}^{\prime}, N_k (= N_j), ... N_{n+1})$ differing from $\mathcal{P}$ in the set of nodes $(N_1^{\prime}, ..., N_{k-1}^{\prime})$. Note that the length of the paths $\mathcal{P}$ and $\mathcal{P}^{\prime}$ can be different.
    \item [6.]  The challenger gives $Pkt_1$ to the adversary. Then, if $b=0$, it gives also packet $Pkt_2$, otherwise, if $b=1$, it gives $Pkt_2^{\prime}$.
    \item [7.] The adversary can submit any number of packets to the challenger to have $N_j$ process them according to the protocol. The resulting packets are given to the adversary. 
    \item [8.] The adversary produces guess $b^{\prime}$.
\end{enumerate}

Path-session unlinkability is achieved if any Probabilistic Polynomial Time (PPT) adversary $A$ cannot guess $b^{\prime}=b$ with a probability non-negligibly better than $0.5$.
\end{definition}

\noindent In step 3 of the game, we do not explicitly state whether the adversary provides the key materials for the nodes different from $N_j$. 
If attacker $A$ belongs to the $\mathcal{A}_2$ class of attackers, then $A$ provides the challenger with the key material for the nodes $N_{i, i \neq j}$ to derive the temporary keys from the shared symmetric master keys.
Otherwise, if $A$ belongs to the $\mathcal{A}_1$ attacker class, it can only access the packets on the links between the nodes, and cannot provide the key material for any node in the network.

\item[\textbf{Source-session unlinkability}] means that an adversary cannot determine if two packets sharing the same path except the source and passing through the same honest node $N_j$ belong to the same session.

\begin{definition}[\textbf{Source-session unlinkability}]
\label{def:source-session-unlinkability}

Consider an adversary interacting with a challenger running our privacy-preserving protocol. The source session unlinkability game runs as follows:

\begin{enumerate}
    \item [1-4.] Same as in the path session unlinkability game (Definition \ref{def:path-session-unlinkability}).
    \item [5.] The challenger sets a bit $b$ at random. It creates a first data packet $Pkt_1$ carrying $p_1$ following path $\mathcal{P}$. Then it creates two other packets carrying $p_2$. A first packet ($Pkt_2$) is following path $\mathcal{P}$. A second packet ($Pkt_2^{\prime}$) is following path $\mathcal{P}^{\prime}=(N_0^{\prime}, N_1, ..., N_j, ... N_{n+1})$ that shares the same nodes as $\mathcal{P}$ except the source.
    \item [6-8.] Same as in the path session unlinkability game.
\end{enumerate}

Source-session unlinkability is achieved if any PPT adversary $A$ cannot guess $b^{\prime}=b$ with a probability non-negligibly better than $0.5$.

\end{definition}
\noindent As with the previous definition, we do not explicitly state  whether  the adversary provides the challenger for the network nodes except the honest node $N_j$.
\item[\textbf{Session unlinkability}] can now be formally defined as:

\begin{definition}[\textbf{Session unlinkability}]
\label{def:session-unlinkability}

Session unlinkability is achieved if, for any PPT adversary $A$, both path-session unlinkability and source-session unlinkability are achieved.

\end{definition}

\end{description}

\subsection{Session unlinkability in Ariadne}
\label{sec:gameproof_ariadne-session-unlinkability}

We detail how Ariadne achieves session unlinkability properties. 

\begin{theorem}
\label{def:theorem-A1}

Ariadne achieves session unlinkability against any PPT adversary $A$ belonging to the $\mathcal{A}_1$ attacker class.

\end{theorem}

\begin{proof}
\label{def:proof-A1}

We consider the packets sent by $N_j$ to $N_{j+1}$.
    According to the onion processing procedure, $Pkt_1$ takes the form:
    \begin{equation}
        \label{eq:first-packet}
        \begin{aligned}
            CH | R \oplus \rho_{j,t,Header} | p_1 &\oplus \rho_{j+1,t,Payload}\\
            ... &\oplus \rho_{n+1,t,Payload}\\
        \end{aligned}
    \end{equation}
    Where $CH$ is the common header consisting of $N_{j+1}$'s address and $p_{j+1,t}$, the pointer to the slot occupied by $N_{j+1}$'s routing element.
    The payload contains $p_1$, sequentially encrypted using $\rho_{i,t,Payload}, j+1 \leq i \leq n+1$.
    The header contains the information for the sub-path taken before node $N_j$, encrypted using the pseudo-random string generated using $k^{Enc}_{j, t}$, while information about the path taken after $N_j$ are encrypted using keys $k^{Enc}_{i, t}, j+1 \leq i \leq n+1$.

    We first prove the path-session unlinkability.
    The two potential packets in the game describing the path-session unlinkability property take the form:
    \begin{equation}
        \label{eq:psu-A1}
        \begin{aligned}
            (Pkt_2)\hspace{0.1cm}  &
            \begin{aligned}
                CH | R \oplus \rho_{j,t+1,Header} | p_2 &\oplus \rho_{j+1,t+1,Payload}\\
                ... &\oplus \rho_{n+1,t+1,Payload}\\
            \end{aligned}\\
            (Pkt_2^{\prime})\hspace{0.1cm}  &
            \begin{aligned}
                CH^{\prime} | R^{\prime} \oplus \rho_{j,t+1,Header} | p_2 &\oplus \rho_{j+1,t+1,Payload}\\
                ... &\oplus \rho_{n+1,t+1,Payload}\\
            \end{aligned}
        \end{aligned}
    \end{equation}
    The pointers to the slot occupied by $N_{j+1}$'s routing element $p_{j+1,t+1}$ and $p^{\prime}_{j+1,t+1}$ are different, as well as the routing header information that is protected by $\rho_{j,t+1,Header}$. 
    Thus, the common header field cannot help making a better guess for $b$. 
    Besides, given that $\rho$ is a pseudo-random generator, the header information encrypted using $\rho_{j,t+1,Header}$ in $Pkt_2$ and $Pkt_2^{\prime}$ are indistinguishable.
    
    We now prove the source-session unlinkability. 
    The two packets presented to the adversary in the game describing the source-session unlinkability property take a similar form as the two packets presented in Equation~\ref{eq:psu-A1}. 
    Indeed, in this game, packets $Pkt_2$ and $Pkt_2^{\prime}$ differ in the pointers to the slot occupied by $N_{j+1}$'s routing element $p_{j+1,t+1}$ and $p^{\prime}_{j+1,t+1}$, as well as the routing header information that is protected by $\rho_{j,t+1,Header}$.
    Using a similar reasoning as in the proof of the path-session unlinkability, it is clear that this data cannot be used by the attacker to guess $b$ with a probability non-negligibly better than $0.5$ in a PPT setting.

    As a conclusion, Ariadne achieves both path-session unlinkability and source-session unlinkability against a PPT adversary belonging to the $\mathcal{A}_1$ attacker class.
\end{proof}

The $\mathcal{A}_1$ class is not as strong as a global adversary that can take control of some nodes in the network. 
This adversary belongs to the $\mathcal{A}_2$ class. 
Ariadne only achieves path-session unlinkability against this class of adversaries because a corrupted node on the path can infer whether two packets were sent using cryptographic material generated with the same key pair.

\begin{theorem}\label{def:theorem-A2}

    Ariadne achieves path-session unlinkability against any PPT adversary $A$ belonging to the $\mathcal{A}_2$ class.

\end{theorem}

\begin{proof}\label{def:proof-A2}

    We consider packets received by node $N_{j+1}$ from $N_{j}$.
    According to the onion processing procedure described in Section~\ref{sec:ariadneproto_data_processing}, the first packet takes a similar form as the packet presented in Equation~\ref{eq:first-packet}.

    The path-session unlinkability property can be proven following the same proof as in Theorem~\ref{def:theorem-A1}. 
    Indeed, even if the adversary now has access to the key material used by several nodes in the network, the information about the path taken before node $N_j$ is protected by the fact that $N_j$'s keys are not compromised and that $\rho$ is a pseudo-random generator.

    Thus, Ariadne achieves path-session unlinkability against any PPT adversary $A$ belonging to the $\mathcal{A}_2$ class.
\end{proof}

\section{Comparing Ariadne's privacy properties with the state-of-the-art}
\label{sec:soa_comparison}

\begin{table*}
    \centering
    \footnotesize
    \makeatletter
    \makeatother
    \begin{threeparttable}
        \caption{Comparison of privacy-preserving communication protocols features}
        \label{tab:protocol_comparison}
        \begin{tabular}{l@{\hskip 0.5in}ccccc}
            Scheme name & 
            {\vtop{\hbox{\strut Onion} \hbox{\strut correctness} }} &
            {\vtop{\hbox{\strut Per hop} \hbox{\strut onion} \hbox{\strut integrity} }} &
            {\vtop{\hbox{\strut Source-tail} \hbox{\strut indistin-} \hbox{\strut guishability} }} &
            {\vtop{\hbox{\strut Path-tail} \hbox{\strut indistin-} \hbox{\strut guishability} }} &
            {\vtop{\hbox{\strut Layer} \hbox{\strut unlinkability} }} \\
            
            \midrule
            State of the art & & & & & \\
            
            \midrule
            Sphinx & \LEFTcircle & \Circle & \Circle & \Circle & \Circle \\
            
            \rowcolor{Gray}
            Enhanced Sphinx & \CIRCLE & \CIRCLE & \CIRCLE & \CIRCLE & \CIRCLE \\

            Tor & \LEFTcircle & \Circle & \Circle & \Circle & \Circle \\

            \cellcolor{Gray}HORNET & \cellcolor{Gray} & \cellcolor{Gray} & \cellcolor{Gray} & \cellcolor{Gray} & \cellcolor{Gray} \\
            \cellcolor{Gray}(Data phase) & \multirow{-2}{*}{\cellcolor{Gray}\LEFTcircle} & \multirow{-2}{*}{\cellcolor{Gray}\Circle} & \multirow{-2}{*}{\cellcolor{Gray}\Circle} & \multirow{-2}{*}{\cellcolor{Gray}\Circle} & \multirow{-2}{*}{\cellcolor{Gray}\CIRCLE} \\

            TARANET & & & & & \\
            (Data phase) & \multirow{-2}{*}{\CIRCLE} & \multirow{-2}{*}{\CIRCLE} & \multirow{-2}{*}{\Circle} & \multirow{-2}{*}{\Circle} & \multirow{-2}{*}{\CIRCLE} \\

            \midrule
            Ariadne & & & & & \\
            
            \midrule
            \textit{- Using Routing} & & & & & \\
            \textit{information vector} & \multirow{-2}{*}{\CIRCLE} & \multirow{-2}{*}{\CIRCLE} & \multirow{-2}{*}{\Circle} & \multirow{-2}{*}{\LEFTcircle} & \multirow{-2}{*}{\LEFTcircle} \\

            \cellcolor{Gray}\textit{- Without Routing} & \cellcolor{Gray}& \cellcolor{Gray}& \cellcolor{Gray}& \cellcolor{Gray}& \cellcolor{Gray}\\
            \cellcolor{Gray}\textit{information vector} & \multirow{-2}{*}{\cellcolor{Gray}\CIRCLE} & \multirow{-2}{*}{\cellcolor{Gray}\CIRCLE} & \multirow{-2}{*}{\cellcolor{Gray}\Circle} & \multirow{-2}{*}{\cellcolor{Gray}\CIRCLE} & \multirow{-2}{*}{\cellcolor{Gray}\CIRCLE} \\
        \end{tabular}

        \begin{tablenotes}
            \item \hfil$\CIRCLE=\text{provides property}$; $\LEFTCIRCLE=\text{partially provides property}$;
            $\text{\Circle}=\text{does not provide property}$
        \end{tablenotes}    
    \end{threeparttable}
\end{table*}

To assess Ariadne's ability to behave as a secure onion routing protocol and present how it improves the state of the art, we compare it with five protocols from the state of the art, namely: Sphinx in its initial version presented in ~\cite{danezis2009sphinx}, Sphinx's enhanced version incorporating modifications from~\cite{Beato2016} and \cite{Kuhn2020}, Tor~\cite{dingledine2004tor}, HORNET~\cite{chen2015hornet} and TARANET~\cite{chen2018taranet}.
We assess Ariadne's properties for two versions of the protocol: 
\emph{i}) Ariadne with Routing Information Vector (cf. Section~\ref{sec:ariadneproto_pathencoding}); and \emph{ii}) Ariadne without Routing Information Vector, using Sphinx's source routing data structure.
This comparison is made according to the secure onion routing properties that we presented and refined earlier in this section. 
We use some security considerations presented in~\cite{Kuhn2020}, in particular regarding Tor, HORNET and TARANET. 
Table~\ref{tab:protocol_comparison} presents the result of this assesment.

All low latency protocols (\textit{i.e.} Tor, HORNET, TARANET and Ariadne) in the comparison are unable to ensure Source-tail indistinguishability. 
The reason is that a corrupted intermediate node can easily determine that two packets generated using the same symmetric key material have been generated by the same source. 
We have already presented the reason why Ariadne does not enforce this property in Section~\ref{sec:gameproof_ariadne-session-unlinkability}.
In Tor, source-tail indistinguishability is not ensured because the packet is processed using cryptographic material referenced by a \textit{Cell ID}. 
This Cell ID is set during the circuit construction procedure, and is specific to a path between a source and destination.
Thus, a corrupted intermediate Tor node can determine that packets using a same Cell ID have taken the same path.
Besides, if the traffic sent between a source and destination is split in multiple subflows, then the Cell ID allows an attacker to correlate packets belonging to a same subflow, which is not sufficient to enforce source-tail indistinguishability.
For the same reason, Tor is not able to ensure path-tail indistinguishability.
In HORNET and TARANET, during the setup phase the source node retrieves a set of \textit{forwarding segments} from the intermediate nodes. 
Those forwarding segments encode the routing instruction and cypher them using a secret key specific to each node. 
In the data phase, for both HORNET and TARANET, an attacker observing the traffic on the links between communicating nodes can observe those forwarding segments in clear in the protocol header.
This is an attack vector by which the attacker can determine that two packets followed the same path, which prevents HORNET and TARANET from ensuring Path-tail indistinguishability.

Ariadne is the only protocol to ensure Path-tail indistinguishability, thanks to the use of the key reference mechanism presented in Section~\ref{sec:ariadneproto_keyderivation}. 
Yet, the use of the routing information vector prevents Ariadne from perfectly ensuring this property, because of the slot position index carried in the common header to reference the routing information vector slot a given node needs to process to retrieve the routing information it needs. 
Indeed, when comparing incoming and outgoing packets, packets carrying the same slot position index cannot belong to the same flow, which gives an adversary an advantage while trying to correlate packest belonging to a same flow.
This leak explains why we have mentionned that Ariadne partially fulfills path-tail undistiguishability in Table~\ref{tab:protocol_comparison}.
The second version of Ariadne, which uses our key referencing method with the classical Sphinx source routing data structure, covers the leak introduced by the slot index, at the expense of packet processing efficiency.

To summarize, Ariadne offers a better privacy protection compared with other low latency privacy-preserving communication protocol. 
In particular, its key referencing mechanism allows Ariadne to be the only protocol in our benchmark to enforce path-tail indistinguishability. 
Due to their use of symmetric key encryption, no low latency privacy-preserving communication protocol is able to ensure source-tail indistinguishability. 
This limitation is inherent to the performance trade-off made to operate with a low latency. 

\section{Evaluation}
\label{sec:evaluation}

\begin{table}[t]
    \centering
    \footnotesize
    \makeatletter
    \makeatother
    \begin{tabularx}
        {\textwidth}
        {l@{\hskip 0.2in}c@{\hskip 0.2in}X} 
        
        Rust crate & Version & Usage \\
        
        \midrule
        \verb|pnet| & 0.31.0 & Network packet creation and processing operations \\ 
        
        \rowcolor{Gray}
        \verb|blake3| & 1.3.1 & Verifying Ariadne packet integrity \\ 
     
        \verb|rand| & 0.8.5 & Random bytes generation using both crypto-grade and fast methods \\ 
     
        \rowcolor{Gray}
        \verb|permutation_iterator| & 0.1.2 & Crypto-grade permutations using Feistel networks \\ 
     
        \verb|fastrand| & 1.7.0 & Fast random permutation \\
     
        \rowcolor{Gray}
        \verb|sha2| & 0.10.2 & Key derivation operations \\
    
        \verb|hkdf| & 0.12.3 & Key derivation operations \\ 
    
        \rowcolor{Gray}
        \verb|x25519-dalek-ng| & git master & Elliptic curve Diffie Hellman key exchange operations based on Curve25519 \\
    
    \end{tabularx}
    \caption{External Rust crates used in Ariadne's implementation}
    \label{tab:Rust-crates}
\end{table}

We implemented Ariadne and its complementary setup protocol (cf. Appendix~\ref{sec:ariadneproto_desc_setup}) using the Rust language, to evaluate the feasibility of our approach. 
We want to determine whether we would be able to generate and process Ariadne traffic at an appropriate speed and provide performance comparable, if not better, to other approaches in real world scenarios. 

\begin{figure*}[t]
    \centering
    \begin{subfigure}[t]{\figsize\textwidth}
        \includegraphics[width=1.15\linewidth]{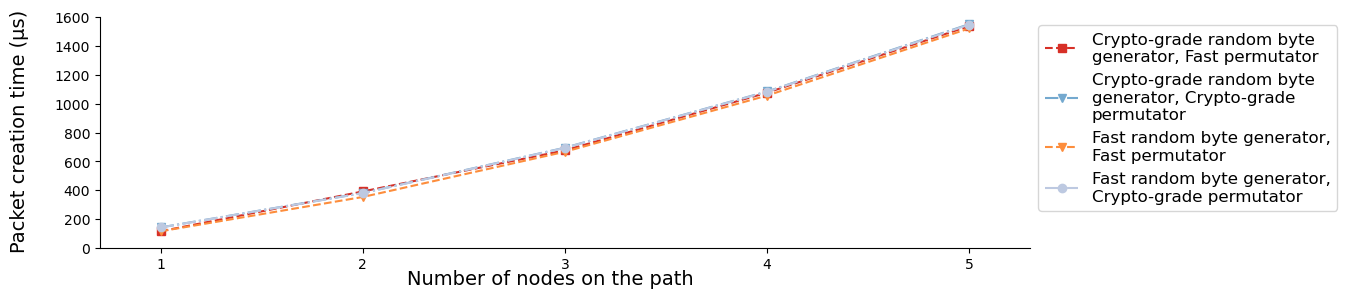}
        \captionsetup{justification=centering}
        \caption{Setup packet generation time depending on\\the number of path nodes}
        \label{fig:sub-Public-key-packet-generation-time}
    \end{subfigure}
    \medskip
    \begin{subfigure}[t]{\figsize\textwidth}
        \includegraphics[width=1.15\linewidth]{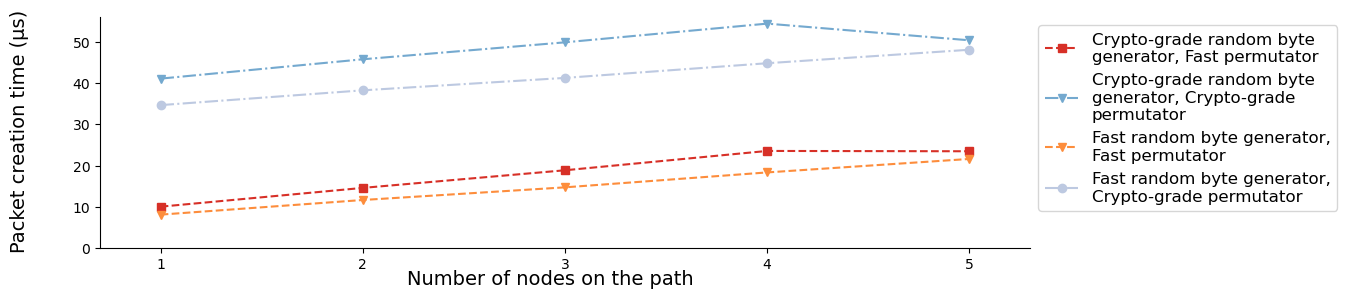}
        \captionsetup{justification=centering}  
        \caption{Data packet generation time depending on\\the number of path nodes}
        \label{fig:sub-Shared-key-packet-generation-time}
    \end{subfigure}
    \medskip
    \begin{subfigure}[t]{\figsize\textwidth}
        \includegraphics[width=1.15\linewidth]{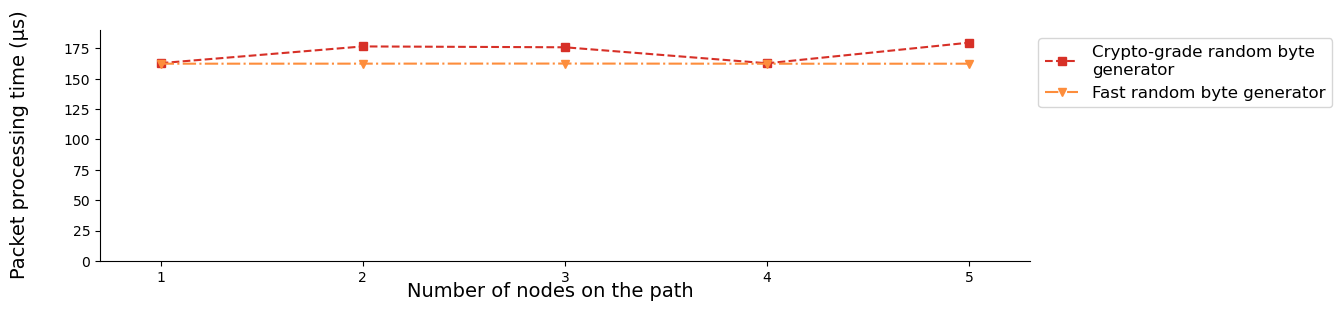}
        \captionsetup{justification=centering}
        \caption{Setup packet processing time at an intermediate node\\depending on the number of path nodes}
        \label{fig:sub-Public-key-packet-processing-time}
    \end{subfigure}
    \medskip
    \begin{subfigure}[t]{\figsize\textwidth}
        \includegraphics[width=1.15\linewidth]{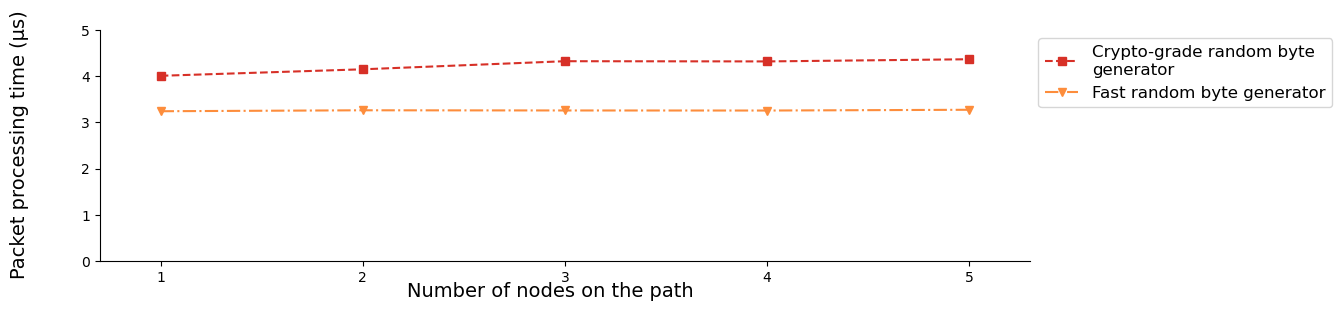}
        \captionsetup{justification=centering}  
        \caption{Data packet processing time at an intermediate node\\depending on the number of path nodes}
        \label{fig:sub-Shared-key-packet-processing-time}
    \end{subfigure}
    \caption{Packet creation and processing time depending on the number of path nodes}
    \label{fig:Packet-creation-processing-time}
\end{figure*}

\subsection{Implementing Ariadne}
\label{sec:evaluation_implementation}

In our implementation, we chose not to use any advanced dataplane packet processing acceleration software, to have an idea of the performance that would be reachable by an end node. 
We have used the Rust programming language for this work given its system programming orientation and its memory safety properties. 
Our implementation uses libraries from several external Rust crates, listed in Table~\ref{tab:Rust-crates}.

Ariadne packets take the form of IPv6 packets carrying an experimental routing extension header. 
This header's design follows the approach of the IPv6 Segment Routing Header specified by the IETF in RFC 8754~\cite{rfc8754}. 
It is identified by a Next header value set to 253 in the experimental range specified in RFC 4727~\cite{RFC4727}.

The Ariadne packet extension header contains an 8 bytes long common header and a routing element vector containing 5 elements which are 36 bytes long.
In those routing elements, the next element address is 16 bytes long and the packet's MAC is also 16 bytes long.
In total, the Ariadne packet extension header is 188 bytes long.
The payload is padded, so that the IPv6 packets carrying our experimental extension header fit the typical Ethernet MTU size of 1500 bytes.

The setup packet extension header contains a common header which is 8 bytes long, a group element which is 32 bytes long and a routing element vector containing 5 elements which are 36 bytes long.
In total, the setup packet extension header is 220 bytes long. 
As in Ariadne packets, the payload is padded to fit the typical Ethernet MTU size of 1500 bytes.

Both setup and Ariadne packets are created at the source node and processed at intermediate nodes following the procedures described in Section~\ref{sec:ariadneproto_desc}, with a small difference: 
in our formal description of the setup onion routing protocol, we use modular exponentiation in the non-interactive Diffie-Hellman key exchange, while in our implementation we use elliptic curve cryptography.
Ariadne's packet operations are transposable to elliptic curve cryptography without harming our protocol's security.

\subsection{Packet creation and processing performance}
\label{sec:evaluation_performance}

We have measured the performance of our implementation using the \verb|criterion| benchmark tool in its version 0.3.5, as this tool is distributed with the Rust toolchain.
We have developed a set of benchmarks to evaluate the time taken by a node to generate and process a set of Ariadne and setup protocol packets. 
In our measurements we have aimed at characterizing the packet creation and processing performance according to the number of nodes on the path (i.e., a linear topology).
In the Ariadne packet processing time, we have considered the pattern matching lookup time, the new key generation time, and the pattern encryption time.
Besides, we also measured the penalty associated with the use of cryptographically secure random byte generators and permutators by comparing them to their performance-oriented counterparts. 

We have run our benchmark on a single server equipped with an Intel Xeon Platinum 8164 processor, clocked at 2 GHz, with 2 sockets, 52 logical cores per socket, and 1.5 TB of RAM. 
Following the \verb|criterion| methodology, we evaluated the creation and the processing of Ariadne packets using one single thread. 
Firstly, performing a warmup of the evaluated parameter for 3 seconds, then collecting 100 samples of the execution time over a period of 5 seconds.
The results of our measurements are presented in Figure~\ref{fig:Packet-creation-processing-time}.
We plot the average running time for the evaluated functions and omit the standard deviation since it is remarkably small across all our measures. 

\noindent \textbf{Symmetric key cryptography performance gain:} The comparison of the Ariadne and setup packet creation and processing time shows a clear benefit associated with the use of symmetric cryptography primitives. 
For instance, if we consider the measures we obtained with 5 path nodes, the source generates a setup packet in 1550.8 $\mu$s (cf. Figure~\ref{fig:sub-Public-key-packet-generation-time}) while it generates an Ariadne packet in 50.353 $\mu$s (cf. Figure~\ref{fig:sub-Shared-key-packet-generation-time}), \textit{i.e.} $30.8$ times faster. 
On average, setup packets are processed in 179.61 $\mu$s (cf. Figure~\ref{fig:sub-Public-key-packet-processing-time}) while Ariadne packets are processed in 4.367 $\mu$s (cf. Figure~\ref{fig:sub-Shared-key-packet-processing-time}), \textit{i.e.} $41.1$ times faster.

\noindent \textbf{Influence of the number of nodes on the path:} Our measures show that the packet creation time depends heavily on the number of nodes on the path. For instance, adding an intermediate fifth node to a path consisting of 4 nodes adds a 466.7 $\mu$s penalty for setup packets (cf. Figure~\ref{fig:sub-Public-key-packet-generation-time}).
However, the number of nodes on the path has little impact on both the setup and the data packet processing performance at intermediate nodes (cf. Figure~\ref{fig:sub-Public-key-packet-processing-time} and Figure~\ref{fig:sub-Shared-key-packet-processing-time}).
This result is interesting because it shows that, in Ariadne, the penalty associated with the use of a longer path is carried by the source nodes, with no impact on intermediate routers.

\noindent \textbf{Throughput:} We also we measured the throughput that a source node can achieve creating Ariadne packets, as well as the throughput an intermediate node can achieve in forwarding these packets. 
On a source node, with a path consisting of 5 nodes, a single thread can generate 28 Gbps of Ariadne traffic, while the intermediate nodes can process 327.5 Gbps of Ariadne traffic. 

We have not yet implemented HORNET or Tor to run a proper benchmark comparison on our platform, yet we can compare the throughput figures we have obtained with the results obtained by HORNET and presented in Section 6.2 of \cite{chen2015hornet}. 
In this paper, the authors show that HORNET can process up to 93.5 Gbps of data traffic on a software router. 
The results we obtained are in the same range, with the addition that Ariadne provides better protection against traffic correlation or Garbage attacks.

\section{Research agenda}
\label{sec:ResearchAgenda}

In the near future, we plan to work on privacy-preserving control plane protocol and public key infrastructure to complement Ariadne and build a full privacy-enhanced network layer. 
In that regard, the work that we have already done on Ariadne is a first step towards the design of a complete privacy-preserving networking stack.

\subsection{Benchmarking and deploying Ariadne}
\label{sec:sub-benchmarking}

In future works, we would like to compare our Ariadne implementation with HORNET or Tor to better assess the merits of our solution, or the penalty associated with the additional protection offered by Ariadne compared with state of the art protocols. 

We would also like to address the challenges that Ariadne would face in real world deployments. Indeed, even if IPv6 is more and more deployed, some destinations on the Internet still need to be reached using IPv4.
Besides, researchers in the Internet measurement community have shown that IPv6 extension headers can be filtered without notice or apparent reason~\cite{leas2022james}. 
To cope with those limitations, we might need to adapt Ariadne to use it together with a transport layer encapsulation mechanism such as Geneve~\cite{rfc8926} or LISP~\cite{rfc9300}.
In such a scenario, Ariadne would resemble Tor~\cite{dingledine2004tor} or Apple Private Relay~\cite{Apple2021}. 

\subsection{Anonymous Network Control Plane}
\label{sec:sub-AnonControlPlane}

The privacy of the communicating parties might be weakened by data leakages occurring while the source gathers routing information or retrieves public keys for intermediate nodes.
Work is needed on how network nodes using Ariadne can privately compute paths to any destination and retrieve associated cryptographic material. 

In the state of the art, three approaches are used to prevent information leakage related to route computation or public key retrieval operations: \emph{(i)} the source node retrieves a full view of the network topology to compute the route by itself, with the complete list of the nodes' public keys; 
\emph{(ii)} the source retrieves information about a subset of the network while hiding the network area it is interested in; 
\emph{(iii)} the route computation is delegated to a third party and privacy-preserving processing methods are used. 
All of these methods present pros and cons, which are worth to be explored in when using Ariadne as a data-plane.

\section{Conclusion}
\label{sec:conclusion}

To address the growing privacy concerns of online service users, we propose Ariadne, a privacy-preserving network layer protocol which ensures that no information can be used to link a packet's source and destination, and that packets cannot be associated together as belonging to the same flow.
We use source routing to avoid relying on trusted third parties. 
Ariadne has been designed to use only symmetric key cryptography in the creation and relay of the packets.

Ariadne uses two innovative techniques to preserve privacy.
First, the source routing information is encoded in a sequentially encrypted, fixed size vector in which the various routing information elements are pseudo-randomly permuted to prevent any attacker from learning topological information about the packet's path in the network.
Second, the temporary symmetric keys used to encrypt the packets are identified using a mutually known encrypted pattern rather than with an explicit reference.

Ariadne achieves \emph{session unlinkability} against a global passive adversary, and \emph{path-session unlinkability} against a stronger global passive adversary able to corrupt some nodes on the path. 
Our first implementation of Ariadne shows the protocol's ability relay private network traffic at an appropriate speed in real world scenarios.

\bibliography{ariadne-preprint}

\begin{appendices}
    
\section{A possible path setup and key exchange extension to Ariadne}
\label{sec:ariadneproto_desc_setup}
    
In this article, we made the assumption that the source node $N_0$ has a shared master symmetric key $k_i$ with each node $N_i$ on the path to a destination $N_{n+1}$ so it can send anonymous network traffic along this path. 
Here, we present an onion routing protocol that can be used to set up the path from the source to the destination, \textit{i.e.}, to exchange master symmetric keys between the source and each node on the path initially leveraging to their public-private keys. 
This setup protocol is based on the Sphinx protocol~\cite{danezis2009sphinx},modified to \emph{(i)} use our routing element vector rather than the Last In-First Out data structure used in Sphinx and \emph{(ii)} add a hop-by- hop integrity check to prevent adversaries from performing a Garbage attack.
    
\subsection{Additional notation}
\label{sec:ariadneproto_setup_notations}

In addition to the notation already introduced in Section~\ref{sec:ariadneproto_data_notations}, we need to introduce other elements in order to formally describe the setup protocol.
    
The setup protocol uses a \emph{setup routing elements vector} to convey the anonymized source-routed path information. We denote $R^s$ this routing vector consisting in elements of length $U^s$ bytes.\footnote{The superscript $^s$ is used to distinguish elements that are specific to the setup protocol.}
The setup routing element stored in this vector used by node $N_i$ is denoted by $r^s_i$.
For $i$ ranging from $1$ to $n$, this routing element is the concatenation of $Addr_{i+1}$, the address of $N_i$'s successor on the path, $p^s_{i+1}$, a pointer giving the position of the next routing element in the routing vector, and $\gamma^s_{i}$, the result of a MAC computed over the whole packet and header. 
In $r^s_{n+1}$, as there is no successor node, the address is $N_{n+1}$'s own address and the pointer is pointing to the actual slot in the routing element vector. This slot starts at byte indexed $P^s_i$.
    
Given that $l_{pmax}$ denotes the maximum number of relays that can be accommodated by the routing element vector, the setup routing element vector's length is $L_{smax} = l_{pmax} \times U^s$ bytes.
    
The setup onion routing protocol we are defining is using public key cryptography in its operations. Let us define the following cryptographic elements for the protocol:
    
\begin{itemize}
    \item $G$ is a prime order cyclic group satisfying the decisional Diffie-Hellman assumption~\cite{Boneh1998}. $G^*$ is the set of non-identity elements of $G$, and $g$ is a generator for this group.
    \item $x_i$ and respectively $y_i = g^{x_i}$ are the private and public keys used by a node $N_i$ in the setup protocol.
    \item $x_t$ and respectively $y_t = g^{x_t}$ are the private and public keys picked by the source node S before sending a setup packet. This key pair is used only for this packet.
    \item $\alpha_i$ is a group element sent by the source $N_0$ to a node $N_i$ so that $N_i$ can compute its shared key with $N_0$. This group element is derived from the packet's public key $y_t$.
    \item $b_i$ is a blinding factor used by a node $N_i$ to derive the group element $\alpha_{i+1}$ to send to its successor on the path during the setup phase. We use a hash function $h_{b}: G^* \times G^* \rightarrow \mathbb{Z}^*_{q}$ to compute those blinding factors.
    \item  $\rho^s: \{0,1\}^\kappa \rightarrow \{0,1\}^{|L_{smax}+U^s|}$ is a pseudo-random generator for the setup protocol, keyed by $k^{Enc}_{i, 0}$ to generate a string of length $|L_{smax}+U^s|$ bits.
    \item  $\mu^s: \{0,1\}^\kappa \times \{0,1\}^{|L_{smax}+U^s|} \rightarrow \{0,1\}^\kappa$ is a Message Authentication Code (MAC) function over the whole packet. It is keyed by $k^{Mac}_{i, 0}$ to generate a MAC $\gamma^s$ of length $\kappa$ bytes.
\end{itemize}
    
Following the same formalism as the Ariadne description, we define the setup protocol as a key generation step and an onion creation step used by the source to create the setup packet (see Section~\ref{sec:ariadneproto_setup_creation}) in addition to a packet processing procedure used by relay nodes along the path (see Section~\ref{sec:ariadneproto_setup_processing}). 

\subsection{Setup packet creation}
\label{sec:ariadneproto_setup_creation}

\noindent \textbf{Key generation algorithm:} 
~\\
To send a setup packet, the source node \textbf{$N_0$} first selects a temporary private key $x_t$ and compute the associated public key $y_t$ using $G$'s generator $g$. $N_0$ will use this key pair to negotiate symmetric keys with the nodes on the path. 
As this key is temporary, it is hard for the other nodes to determine $N_0$'s identity from it.
Then, \textbf{$N_0$} gathers the list of public keys $y_0$, $y_1$, ... $y_{n}$ and $y_{n+1}$ of the nodes constituting the path to the destination.
With those public keys, it computes the elements to be used during the routing vector creation procedure, namely: the group elements $\alpha$, the shared symmetric keys $k$, and the blinding factors $b$:
\begin{equation}
    \begin{aligned}
        \text{Node \textbf{$N_1$}:}\quad &
        \begin{aligned}
            &\alpha_1=g^{x_t}\\
            &k_1=y_1^{x_t}=\alpha_1^{x_1} \\
            &b_1=h_b(\alpha_1, k_1)
        \end{aligned}
        \\
        .........\\
        \text{Node \textbf{$N_i$}:}\quad &
        \begin{aligned}
            &\alpha_i=g^{x_t b_1...b_{i-1}}=\alpha_{i-1}^{b_{i-1}}\\
            &k_i=y_i^{x_t b_1...b_{i-1}}=\alpha_i^{x_i}\\
            &b_i=h_b(\alpha_i, k_i)
        \end{aligned}
    \end{aligned}
\end{equation}
At last, for each node, $N_0$ will derive the MAC key $k^{Mac}_{i, 0}$, the encryption key $k^{Enc}_{i, 0}$ and the pseudo random string used to encrypt the packet using $\rho^s$:
\begin{equation}
    \begin{aligned}
        &(k^{Enc}_{i, 0}; k^{Mac}_{i, 0}) = \delta(k_i, 0)\\
        &\rho^s_{i,r} = \rho^s(k^{Enc}_{i, 0})_{[0; |U^s-1|]}\\
        &\rho^s_{i,Header} = \rho^s(k^{Enc}_{i, 0})_{[|U^s|; |U^s + L_{smax}|]}\\
        &\rho^s_{i,Packet} = \rho^s(k^{Enc}_{i, 0})_{[|U^s|; |L_{full}-1|]}
    \end{aligned}
\end{equation}

\noindent \textbf{Onion creation algorithm:} 
~\\
After the key generation procedure, the source \textbf{$N_0$} initializes a routing elements vector $R^s$ consisting of $l_{pmax}$ slots of size $U^s$ bytes. 
Then it computes a pseudo-random permutation over the $l_{pmax}$ slots, to determine the slot positions $p^s_1$, $p^s_2$,... $p^s_n$ and $p^s_{n+1}$ at which it will place the routing information elements that will be used by each node. $N_0$ then builds a filler byte string using the $r^{s*}_i$ routing elements.
Starting from $N_1$'s information until $N_{n+1}$'s, $N_0$ build $R^s$ from the initial random byte string by doing the following operations:
\begin{equation}
    \begin{aligned}
        &r^{s*}_i = Addr_{i+1} | p^s_{i+1} | 0_\kappa\\
        &R^s = R^s_{[0; |P^s_i - 1|]} | r^{s*}_i | R^s_{[|P^s_i + U^s|; |L_{smax}-1|]}\\
        &R^s = R^s \oplus \rho^s_{i,Header}\\
    \end{aligned}
\end{equation}
Then, $N_0$ can start building the routing elements vector $R^s$. As for $R$, this construction is made backwards, starting from the destination $N_{n+1}$.
$N_0$ concatenates the $R^s$ filler with the packet's payload and a potential padding to reach a size of $L_{Payload}$.
    
Then, for each node on the path, starting with node $N_{n+1}$ down to node $N_1$, $N_0$ follows the same onion layer creation procedure, described here for node $N_i$. 
First, $N_0$ encrypts the packet using the $\rho^s_{i,Packet}$ byte string. 
It computes the MAC $\gamma^s_i$ of the resulting packet using $\mu^s$ keyed by $k^{Mac}_{i, 0}$.
The result, $\gamma^s_i$, is used with $N_{i+1}$'s address and routing element slot pointer $p^s_{i+1}$ to create the routing element $r^s_i$, which is encrypted using $\rho^s_{i,r}$. 
The resulting encrypted routing element is placed at the slot designated by $p^s_i$ in the routing element vector $R^s$.  
\begin{equation}
    \begin{aligned}
        &X = X \oplus \rho^s_{i,Packet}\\
        &\gamma^s_i = \mu^s(k^{Mac}_{i, 0}; X)\\
        &r^s_i = Addr_{i+1} | p^s_{i+1} | \gamma^s_i\\
        &X = X_{[0; |P^s_i - 1|]} | r^s_i  \oplus \rho^s_{i,r} | X_{|[P^s_i + U^s|; |L_{full}|]}\\
    \end{aligned}
\end{equation}
Once the routing vector is created, a common header containing $N_1$'s address $Addr_{N_1}$ in clear is prepended, as well as the group element $\alpha_1$ and the pointer $p^s_1$ used to find the next node. The packet is then ready to be sent to $N_1$.
    
\subsection{Setup packet processing}
\label{sec:ariadneproto_setup_processing}

Let's consider the packet processing and relaying operation (following the \textbf{Onion processing algorithm}) performed by node $N_i$.

When $N_i$ receives a setup packet, it first retrieves $\alpha_i$ from the common header and uses its private key $x_i$ to compute the shared key $k_i$ and associated cryptographic material:
\begin{equation}
    \begin{aligned}
        &k_i=\alpha_i^{x_i}\\
        &\delta(k_i, 0) = (k^{Enc}_{i, 0}; k^{Mac}_{i, 0})\\
        &\rho^s(k^{Enc}_{i, 0}) \rightarrow \rho^s_{i,r}; \rho^s_{i,Header}; \rho^s_{i,Packet}\\
    \end{aligned}
\end{equation}
Then, it looks at the routing element pointer in the common header to retrieve the encrypted routing element $r^s_i$. It uses $\rho^s_{i,r}$ to decrypt it and to retrieve the MAC $\gamma^s_i$ for the packet:
\begin{equation}
    \begin{aligned}
        &r^s_i = X_{[|P^s_i|; |P^s_i + U^s - 1|]} \oplus \rho^s_{i,r}\\
        &r^s_i \rightarrow (r^{s*}_i; \gamma^s_i)\\
    \end{aligned}
\end{equation}
It replaces $r^s_i$ with $r^{s*}_i$ in the routing element vector $R^s$ to compute the packet's MAC and compare it to the retrieved $\gamma^s_i$.
\begin{equation}
    \begin{aligned}
        &X = X_{[0; |P^s_i-1|]} | r^{s*}_i | X_{[|P^s_i + U^s|; |L_{full}|]}\\
        &\gamma^{s\prime}_i = \mu^s(k^{Mac}_{i, 0}; X)
    \end{aligned}
\end{equation}
If $\gamma^s_i$ and $\gamma^{s\prime}_i$ are equal, then the packet can be decrypted, which blinds the routing information element $r^s_i$ at the same time. The common header is updated with the information extracted from $r^s_i$, and $\alpha_i$ is replaced with a new value $\alpha_{i+1}$: 
\begin{equation}
    \begin{aligned}
        &X = X \oplus \rho^s_{i,Packet}\\ 
        &b_i=h_b(\alpha_i, k_i)\\
        &\alpha_{i+1}=g^{x_t b_i}=\alpha_1^{b_i}\\
    \end{aligned}
\end{equation}
The packet is then sent to the next node, $N_{i+1}$. When $N_{n+1}$ receives the packet, it begins processing it like the previous intermediate nodes. When it finds out the address of the next node in $r^s_{n+1}$ is its own address, then it checks the packet integrity, and if the packet has not been corrupted it proceeds with the decryption of the packet. 
    
By sending a setup protocol packet to node $N_{n+1}$ along the path, the source $N_0$ has negotiated shared keys with each intermediate nodes on the path. Those shared keys can be used by each node to process Ariadne packets. Other key exchange and distribution mechanisms might be used depending on the context in which Ariadne operates.
    
\end{appendices}
    
\end{document}